\documentclass[twocolumn]{aastex63}
\usepackage[mediumspace,mediumqspace,Grey,squaren]{SIunits}
\usepackage{amssymb,mathrsfs,amsmath}
\usepackage{CJK}
\usepackage{lineno}
\graphicspath{{./}{figures/}}

\submitjournal{ApJ}

\shorttitle{Stellar populations in NGC 2210}
\shortauthors{Li et al.}

\begin{document}
\begin{CJK*}{UTF8}{gbsn}

\title{Multiple stellar populations at less evolved stages-III: a possible helium spread in NGC 2210}

\correspondingauthor{Chengyuan Li}
\email{lichengy5@mail.sysu.edu.cn}

\author{Chengyuan Li (李程远)} 
\affil{School of Physics and Astronomy, Sun Yat-sen University, Daxue Road, Zhuhai, 519082, China}
\affil{CSST Science Center for the Guangdong-Hong Kong-Macau Greater Bay Area, Zhuhai, 519082, China}
\author{Xin Ji (纪鑫)}
\affil{Key Laboratory of Optical Astronomy, National Astronomical Observatories, Chinese Academy of Sciences, Beijing 100101, China}
\affil{School of Astronomy and Space Science, University of Chinese Academy of Sciences, Beijing 100049, China}

\author{Long Wang (王龙)}
\affil{School of Physics and Astronomy, Sun Yat-sen University, Daxue Road, Zhuhai, 519082, China}
\affil{CSST Science Center for the Guangdong-Hong Kong-Macau Greater Bay Area, Zhuhai, 519082, China}
\author{Yue Wang (王悦)}
\affil{Key Laboratory of Optical Astronomy, National Astronomical Observatories, Chinese Academy of Sciences, Beijing 100101, China}

\author{Baitian Tang (汤柏添)} 
\affil{School of Physics and Astronomy, Sun Yat-sen University, Daxue Road, Zhuhai, 519082, China}
\affil{CSST Science Center for the Guangdong-Hong Kong-Macau Greater Bay Area, Zhuhai, 519082, China}

\author{Antonino P. Milone}
\affil{Dipartimento di Fisica e Astronomia ``Galileo Galilei'', Univ. di Padova, Vicolo dell'Osservatorio 3, Padova, IT-35122, Italy}
\affil{Istituto Nazionale di Astrofisica - Osservatorio Astronomico di Padova, Vicolo dell'Osservatorio 5, Padova, IT-35122, Italy}
\author{Yujiao Yang (杨玉姣)}
\affil{Key Laboratory of Space Astronomy and Technology, National Astronomical Observatories, Chinese Academy of Sciences, Beijing 100101, China}
\author{Holger Baumgardt}
\affil{School of Mathematics and Physics, The University of Queensland, St. Lucia, QLD 4072, Australia}
\author{Dengkai Jiang (姜登凯)}
\affil{Yunnan Observatories, Chinese Academy of Sciences, 396 Yangfangwang, Guandu District, Kunming 650216, China}
\begin{abstract}
Helium variations are common features of globular clusters (GCs) with multiple stellar populations. All the formation scenarios predict that secondary population stars are enhanced in helium but the exact helium content depends on the polluters. Therefore, searching for helium variations in a star cluster is a straightforward method to understand if it hosts multiple populations or not, and constrain the formation scenario.  Although this topic has been well explored for Galactic GCs, GCs beyond the Milky Way are challenging to study because of their large distances. This work studies the helium distribution of GK-type main sequence dwarfs in an old ($\sim$12.5 Gyr) GC in the Large Magellanic Cloud, NGC 2210, using the deep photometry observed by the {\sl Hubble Space Telescope}. We compare the observed morphology of the MS with that of synthetic populations with different helium distributions. We confirm that NGC 2210 dwarfs have a helium spread, with an internal dispersion of $\delta{Y}\sim$0.06--0.07. The fraction of helium enriched stars depends on the $\delta{Y}$ distribution. A continuous $\delta{Y}$ distribution would indicate that more than half of MS stars are helium enriched ($\sim$55\%). If the $\delta{Y}$ distribution is discrete (bimodal), a fraction of $\sim$30\% enriched stars is able to explain the observed morphology of the MS. We also find that the He-enriched population stars are more centrally concentrated than He-normal stars. 

\end{abstract}


\section{Introduction} \label{S1}
In contrast to young open clusters and star-forming regions, which have been proved to be simple stellar populations \citep[SSPs. e.g.,][]{deSi09a,Brag12a,Brag14a,Ting12a,Kos21a}, most globular clusters (GCs) and some intermediate-age clusters (older than $\geq$2 Gyr) are multiple stellar populations \citep[MPs, e.g.,][]{Carr09a,Milo17a,Nied17a,Li19a}. The MPs are characterized by star-to-star chemical variations in light elements such as C, N, O, Na, Mg, Al \citep{Carr09a,Mari16a,Panc17a,Dias18a}, together with He \citep{Piot07a,Milo18a}. 

The observed chemical pattern of MPs in GCs drives the debate on suitable polluters. Although various scenarios were proposed to explain the presence of MPs, none of these models can reproduce the exact pattern of abundances observed in GCs \citep[see][]{Bast15a}. These models invoke interactive binaries \citep{deMi09a,Jiang14a,Wang20a,Wei20a,Renz22a}, fast-rotating massive stars \citep[FRMS,][]{Krau13a}, asymptotic giant branch (AGB) stars \citep{Derc08a,Dant16a,Calu19a}, very massive stars \citep[$10^2$--$10^3$ $M_{\odot}$,VMS,][]{Vink18a}. Concerning nucleosynthesis, all these models predict polluted stars should be He-enriched as He is the direct product of H-burning. The helium abundance is also used as a proxy for the chemical enrichment in numerical simulations \citep[e.g.,][]{Howa19a}.

Although a helium spread is expected in all GCs with MPs, direct helium measurements ($Y$) are challenging, as most stars in GCs are too cold to exhibit He lines. Because He absorption lines can only be detected among horizontal branch (HB) stars (hotter than $\sim$9,000 K) in GCs\footnote{In addition, HB stars used for He determination should not be hotter than $\sim$11,000 K to avoid the Grundahl jump effect \citep{Grun99a}.}\citep[e.g.,][]{Mari14a,Dupr11a,Dupr13a}. An alternative method is based on the photometric investigation of GC stars. As helium is the second most abundant element in stars, its variation has a notable impact on stellar structure and evolution. Stellar evolutionary theory predicts that He-rich stars will be hotter and brighter than normal stars at each stage, as He-rich stars have a smaller radiative opacity and a higher mean molecular weight than normal stars. In addition, He-rich stars evolve more rapidly than normal stars as they have increased luminosities. At a given age, He-rich stars at the MS turnoff (TO) stage will be less massive, populating a fainter MSTO boundary. If both He-rich and normal stars experience the same mass loss content during their red-giant branch (RGB) stage, they will end up with different masses when evolve into the HB, thus different colors. The current stellar evolutionary model also predicts that the RGB bump (RGBB) lifetime would be shortened for He-enhanced stars \citep{Bono01a,DiCe10a}. Indeed, helium distributions have been studied in some GCs through photometry based on the morphologies of MS \citep[e.g., ][]{Piot07a}, RGB \citep[e.g.,][]{Milo17a}, RGBB \citep[e.g., ][]{Nata11a,Lagi19a}, and HB \citep[e.g., ][]{Jang19a}. Statistical studies based on Galactic GCs have shown that both the maximum internal helium dispersion ($\delta{Y}$), and the fraction of helium-enriched stars positively correlate with the total clusters' masses \citep{Milo17a,Milo18a}. This correlation also applies to extragalactic GCs \citep[][hereafter C19]{Chan19a}.

The method based on the HB could overestimate the internal helium variation if it does not account for the mass loss effect \citep{Tail20a}. Studying the helium distribution of less-evolved stars such as MS stars would be more reliable. This is difficult for extragalactic clusters because their large distance requires ultra-deep photometry. A recent attempt was made by \cite{Li21b}, in which they have studied the helium distribution of MS stars for a 1.7 Gyr-old LMC cluster, NGC 1846. They find that NGC 1846 is consistent with an SSP cluster and that its helium spread, if present, should be smaller than 2\%. Another LMC cluster, NGC 1978, was studied by \cite{Ji22a}, in which they have analyzed the morphology of its RGBB. However, they can only constrain its maximum helium spread to $\delta{Y}\leq0.03$ (3\%) due to the limitation of the image quality. This is consistent with \cite{Milo20a} ($\delta{Y}=0.002\pm0.003$).

This work aims to study the helium distribution of MS dwarfs in an old GC, NGC 2210, in the LMC, which serves as a good comparison to our previously studied younger LMC cluster, NGC 1846 \citep{Li21b}. Since only clusters older than $\sim$2 Gyr are known to harbor MPs \citep{Bast18a}, the old cluster NGC 2210 is expected to have a significant helium spread, both in terms of the maximum internal helium spread, and the fraction of He-enriched stars. This work aims to examine this expectation. We present the data reduction and method designation in Section \ref{S2}, and the main results are in Section \ref{S3}. A discussion about our results is present in Section \ref{S4}. 

\section{Methods and data reduction} \label{S2}
The data used in this work was observed through the Advanced Camera for Surveys (ACS) Wide Field Channel (WFC) of the {\sl Hubble Space Telescope} ({\it HST}), obtained through the Mikulski Archive for Space Telescope (MAST). The program ID is GO-14164 (PI: Sarajedini). NGC 2210 was observed through the F606W and F814W filters, with total exposure times of 4306 s and 6950 s, respectively. We do not use another frame observed through the F336W filter of the WFC3 Ultraviolet and Visual channel in this GO program, because most GCs with MPs have star-to-star nitrogen variations, which will produce a deep NH-absorption line centered at $\sim$3370 \AA. A point-spread-function (PSF) based photometry was applied to all charge-transfer-efficiency corrected frames (the `{\tt \_flc}' and `{\tt \_drc}' frames), using the specific {\it HST} photometric package  {\sc Dolphot2.0} \citep{Dolp11a,Dolp11b,Dolp13a}. Similar to \cite{Li21b}, we filter the raw stellar catalog to remove bad pixels, centrally saturated objects, extended sources, cosmic rays and objects being contaminated by crowding. 

\begin{figure}[!ht]
\includegraphics[width=\columnwidth]{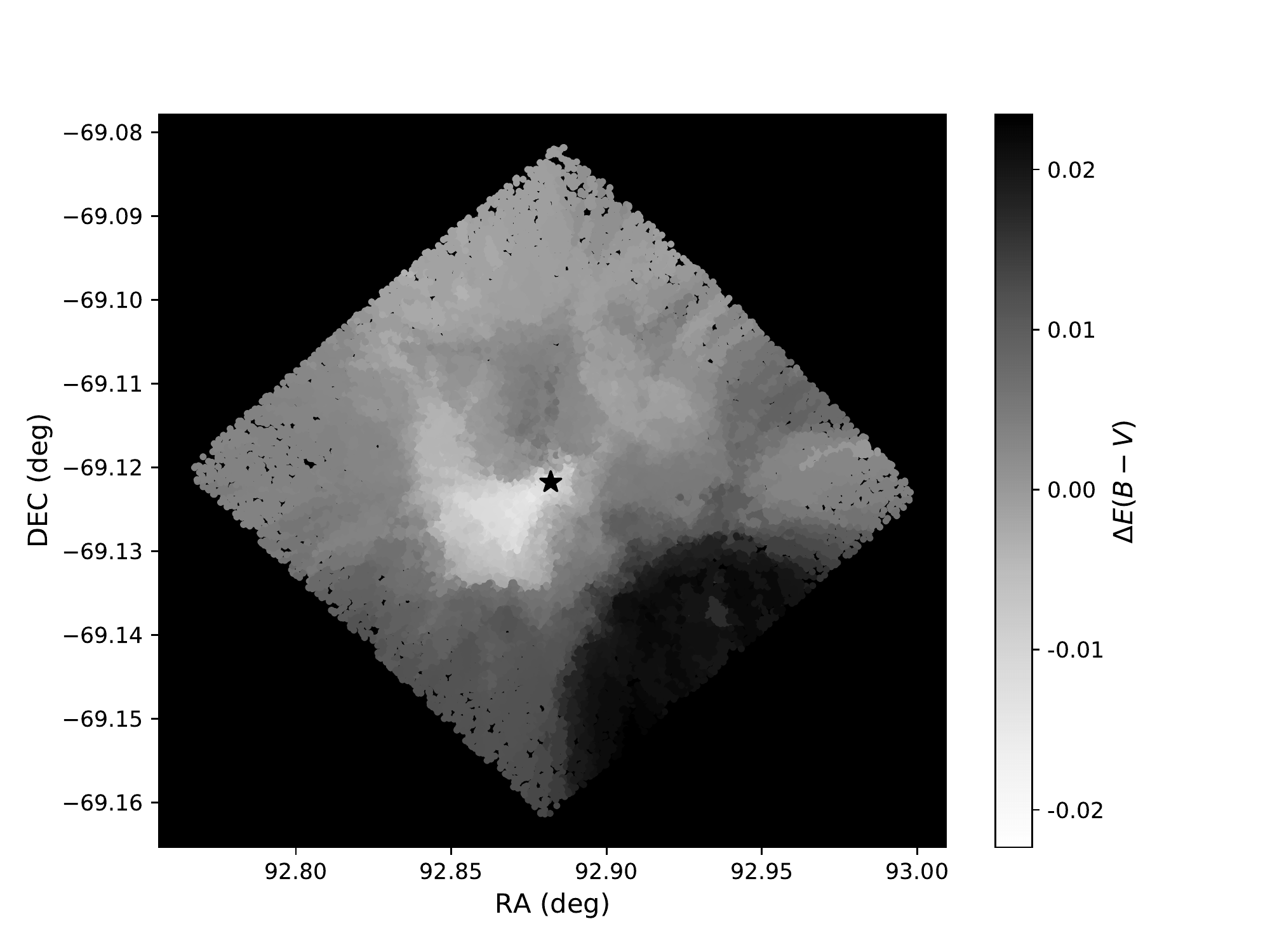}
\caption{The differential reddening ($\Delta{E(B-V)}$) map for all stars in the FoV of the NGC 2210, where $\Delta{E(B-V)}$ is the reddening difference between the referenced star and the median reddening of all stars in the field (color-coded). The standard deviation of the differential reddening is $\sigma_{E(B-V)}$=0.008 mag.}
\label{F1}
\end{figure}

\begin{figure*}[!ht]
\includegraphics[width=2\columnwidth]{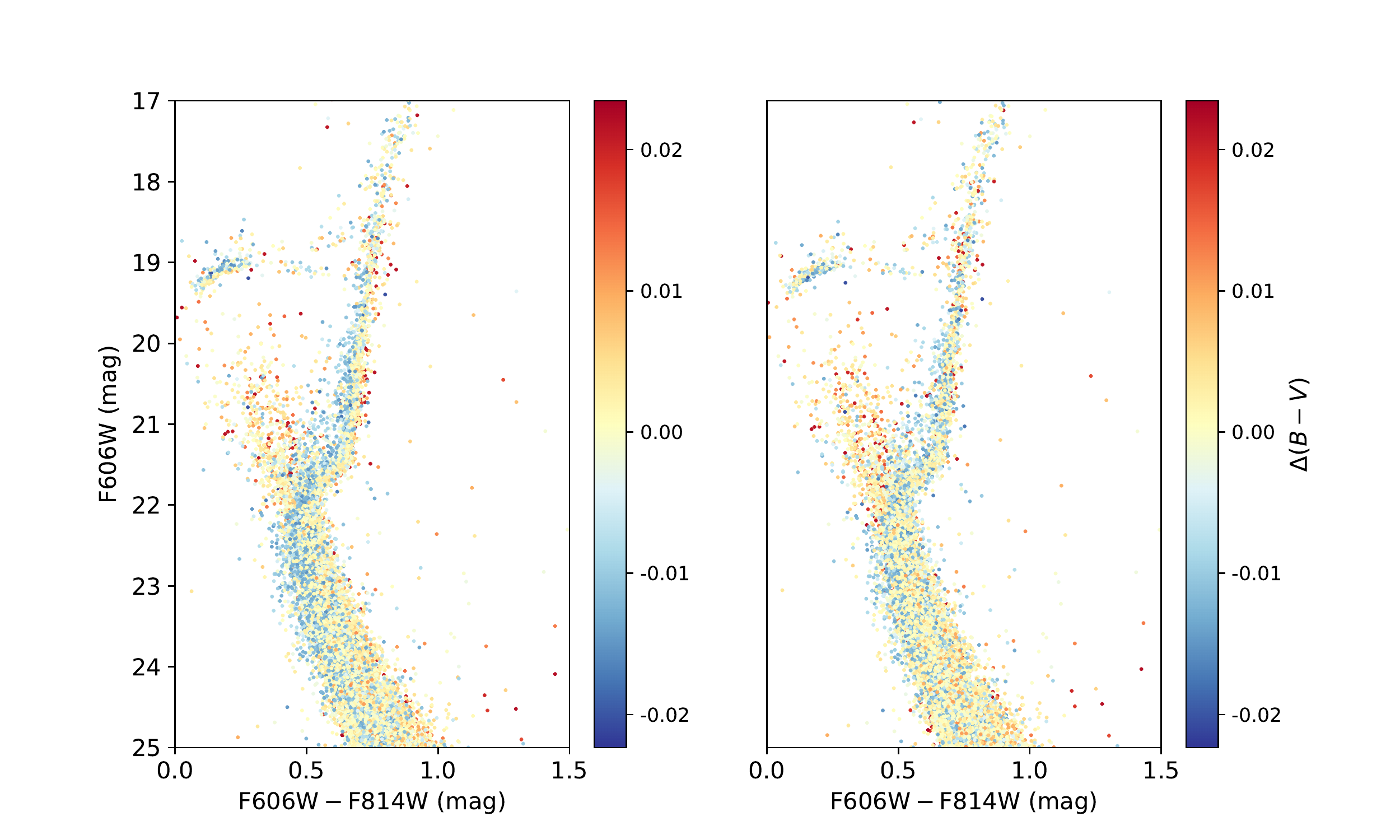}
\caption{Left/right: CMDs of NGC 2210 before/after differential reddening correction.}
\label{F2}
\end{figure*}

The dust distribution in the ACS/WFC field of view (FoV, 202$''\times$202$''$, corresponding to 48.5 pc$\times$48.5 pc at the distance of the LMC) may be inhomogeneous. We have used the method designed in \cite{Milo12a} to minimize the effect caused by the dust inhomogeneity -- the differential reddening. We find a non-negligible differential reddening across the whole FoV for NGC 2210, with a standard deviation of $\sigma_{E(B-V)}$=0.008 mag, which will lead to an average color variation of $\sigma_{({\rm F606W-F814W})}$=0.027 mag. In Fig.\ref{F1}, we exhibit the differential reddening map for all stars observed in the NGC 2210 field. In Fig.\ref{F2} we show the color-magnitude diagrams (CMDs) before/after differential reddening correction. We find that the observed colors of stars positively correlate with their differential reddening, i.e., stars with negative differential reddening are bluer than those with positive differential reddening. We estimate that a residual of differential reddening of $\sigma_{E(B-V)/40}$=0.0002 mag cannot be statistically removed\footnote{ We use the 40 nearest stars surrounding each individual star to correct for differential reddening. To avoid a possible  underestimation of the helium spread, we used RGB rather than MS as the referenced population \citep[see][]{Milo12a}.}. However, the assumption that there is a single referenced ridge line may not be valid if a genuine spatial-dependent helium distribution is present, thus introducing additional uncertainty. In particular, a helium enrichment would mimic a negative differential reddening. Our method for reducing differential reddening may also correct the color shift caused by helium spread, leading to an underestimate of helium spread.

\begin{figure*}[!ht]
\includegraphics[width=2\columnwidth]{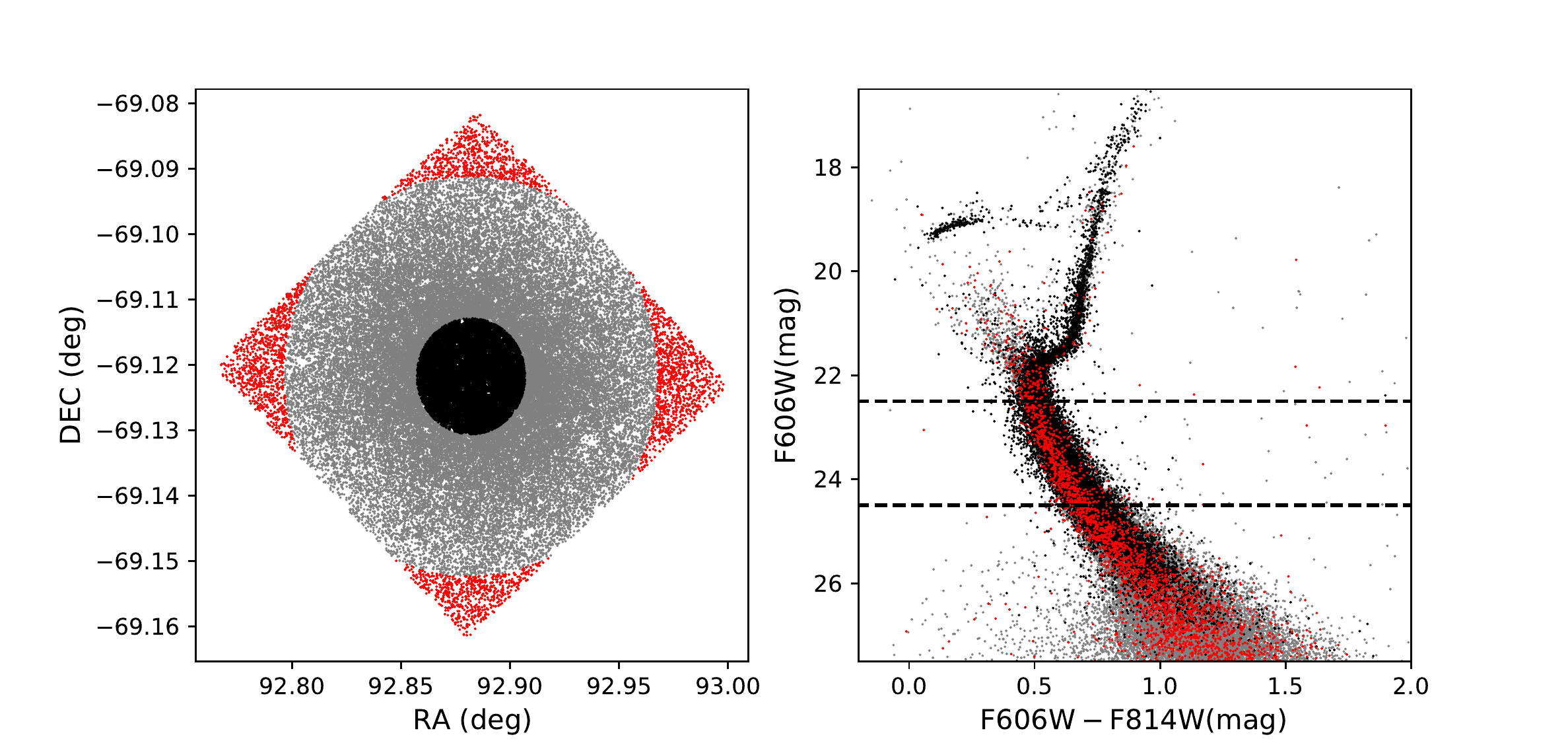}
\caption{Left: spatial distribution of stars in the NGC 2210 field. Right: the CMD of all stars observed in the NGC 2210 field. Black and red dots are selected cluster stars and referenced field stars, respectively. Only stars within the magnitude range defined by dashed lines will be used for analysis.}
\label{F3}
\end{figure*}

Using the world coordinate system (WCS) parameters stored in the header of `{\tt \_drc}' frames, we convert the observed pixel coordinates (X and Y) into the equatorial coordinate system (the $\alpha_{\rm J2000}$ and $\delta_{\rm J2000}$, the right ascension and declination). We directly adopt the center coordinates, $\alpha_{\rm J2000}$=06$^{\rm h}$11$^{\rm m}$31.69$^{\rm s}$, $\delta_{\rm J2000}$=$-69^{\circ}07'18.37''$, as well as the half-mass radius, $r_{\rm h}$=15.9$^{+0.6}_{-0.2}$ arcsec (3.9 pc), for NGC 2210, based on \cite{Ferr19a}. We divide our sample into two parts, stars with a radius from the center, $r\leq32$ arcsec (about 2$r_{\rm h}$) are defined as cluster stars. Stars with $r\geq110$ arcsec (about 7$r_{\rm h}$), which is similar to the tidal radius ($107.1^{+7.6}_{-4.8}$ arcsec, through the King model) determined in \cite{McLa05a}, are defined as referenced field stars. The spatial distribution and the CMD of NGC 2210 stars are presented in Fig.\ref{F3}, with cluster and referenced field stars being color-coded (black and red).

In this work only MS stars within the magnitude range of 22.5 mag$<$F606W$<$24.5 mag are analyzed (stars between the dashed lines in Fig.\ref{F3}), corresponding to a mass range from 0.60$M_{\odot}$ to 0.78$M_{\odot}$. We only select these stars for three reasons: (1) These stars are all located well below the MSTO region. We exclude stars near the TO region because a possible helium spread may complicate the morphology of the TO region (see Section \ref{S1} for an explanation). (2) The photometric uncertainties for these stars are small enough to study broadening caused by possible helium spread. (3) The average completeness of these stars is sufficient for obtaining statistically robust results, which are $\sim$65.8\% for cluster stars and $\sim$82.6\% for referenced field stars (calculated through artificial star test, see below).

\begin{figure*}[!ht]
\includegraphics[width=2\columnwidth]{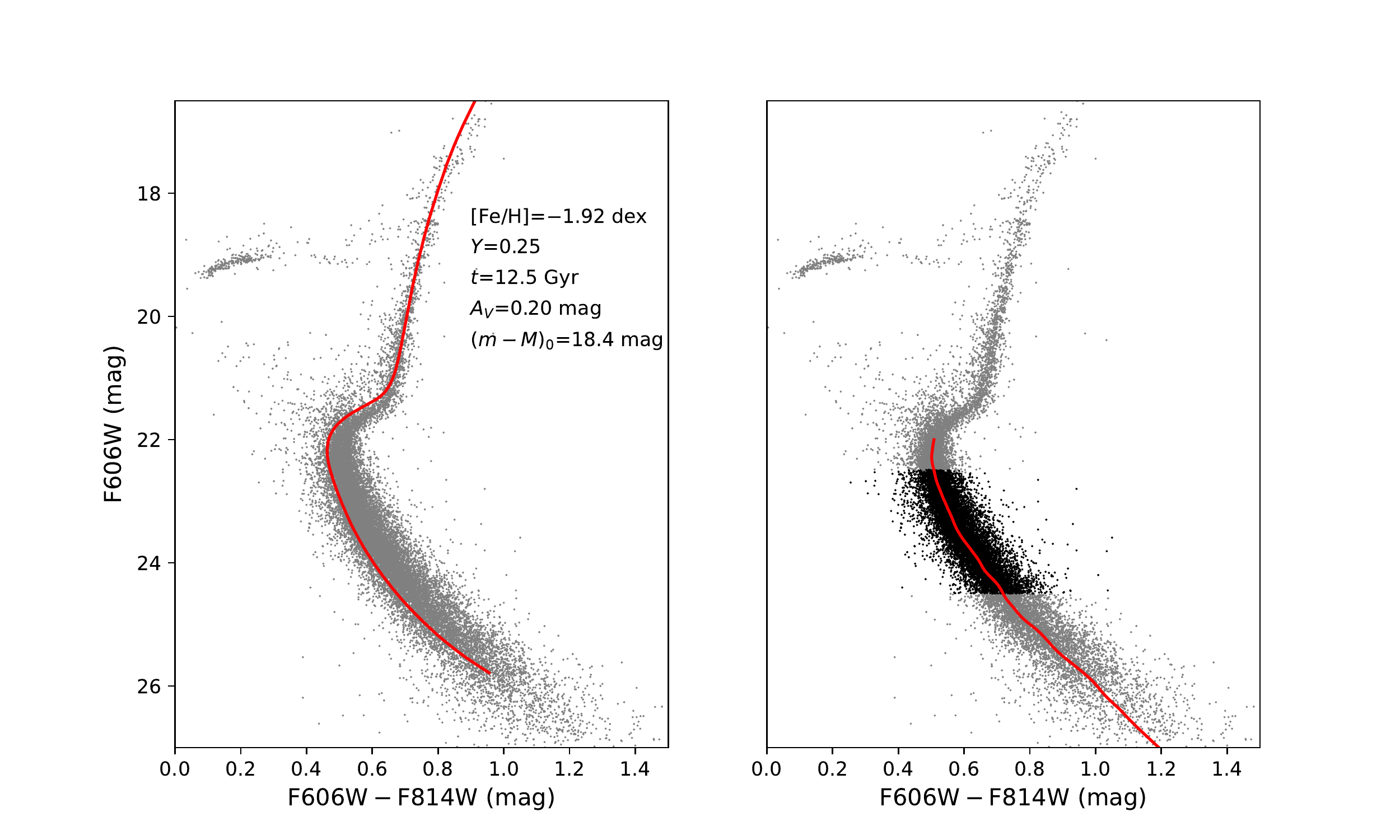}
\caption{Left, the CMD of the NGC 2210 with the best-fitting PGPUC isochrone. The best-fitting parameters are shown in the legend. Right, the same as the left panel, with the calculated MSRL. Black dots are stars used for analysis in this work.}
\label{F4}
\end{figure*}

We used the Princeton-Goddard-PUC (PGPUC) stellar evolutionary code to determine the cluster parameters through isochrone fitting \citep{Valc12a,Valc13a}\footnote{\url http://www2.astro.puc.cl/pgpuc/index.php}. The PGPUC stellar evolution code is focused mainly on the study of low-mass stars (i.e., stars in GCs), which allows users to freely input the values of age, helium abundance ($Y$), global metallicity ($Z$), solar scaled abundance of alpha element ($[\alpha/{\rm Fe}]$) and Mass loss rate ($\eta$) to generate different isochrones. Since the goal of our work is to study the helium distribution of low-mass MS stars in NGC 2210, PGPUC is the most suitable tool for modeling the observation. We determine the best fitting isochrone through visual inspection, which is present in the left panel of Fig.\ref{F4}. The best-fitting parameters are $\log({t/{\rm yr}})$=10.10 ($\sim$12.5 Gyr), $Y$=0.25, [Fe/H]=$-$1.92 dex ($Z$=0.0002 with $Z_{\odot}$=0.0167 in PGPUC model), ($m-M)_0$=18.40 mag (47.86 kpc) and $E(B-V)$=0.06 mag ($A_{V}$=0.20 mag). We have also assumed an enhanced [$\alpha$/Fe]=$+$0.30 dex \citep{Wagn17a} and a default mass loss rate of $\eta$=0.20. The latter is important for stars at post-MS stages (e.g., HB), which does not affect our results since we only concern MS dwarfs. Our best-fitting age, distance modulus and average reddening are comparable to those determined in \cite{Wagn17a} (our age of 12.5 Gyr vs. theirs of 11.63 Gyr$^{+1.80}_{-1.12}$ Gyr; distance modulus of ($m-M)_0$=18.40 mag vs. theirs of ($m-M)_0$=18.523$\pm$0.042 mag; $E(B-V)$=0.06 mag vs. theirs of $E(B-V)$=0.06--0.08 mag), but the adopted metallicity is lower than the spectroscopic study of \cite{Mucc10a} (our metallicity of [Fe/H]=$-$1.92 dex vs. theirs of [Fe/H]=$-$1.65 dex). However, we find even the best-fitting isochrone does not satisfactorily fit the observation, which describes the ridge-line of the RGB but only fits the blue and bright sides of the MS and the SGB. In this work, we want to generate synthetic populations as close to the real observation as possible. We used the MS ridge-line (MSRL) instead of the best-fitting isochrone to model artificial stars. We calculate the MSRL below the TO region using the Gaussian process and iterative trimming (ITGP) based robust regression method developed by \citet[][]{LL20a,LZ21a}\footnote{\url https://github.com/syrte/robustgp/}. The MSRL is shown in the right panel of Fig.\ref{F4}. 

\begin{figure*}[!ht]
\includegraphics[width=2\columnwidth]{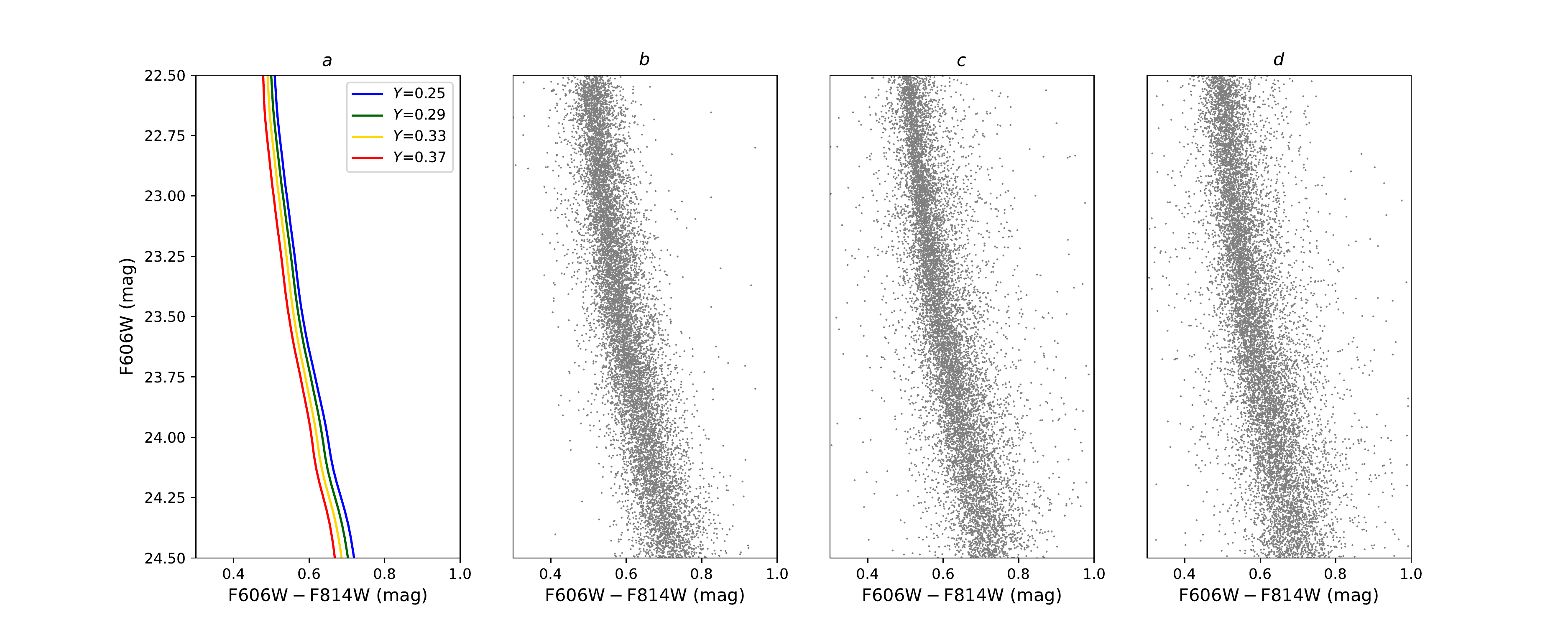}
\caption{Panel $a$: loci of synthetic populations with $Y$=0.25,0.29,0.33,0.37. $b$: The observed MS. $c$: A synthetic SSP with $Y$=0.25. $d$: A synthetic MPs with $Y$ ranges from 0.25 to 0.37.}
\label{F5}
\end{figure*}

Based on the MSRL, we evaluate the effect of He variation through the PGPUC stellar evolution code. We calculate eleven isochrones with He-enrichments of $\Delta{Y}$=0.01--0.12 ($Y$=0.26--0.37, with a step size of 0.01), i.e., a total of twelve isochrones (including the $Y$=0.25 isochrone). The color deviations for each isochrone to the standard isochrone ($Y$=0.25), $\Delta($F606W$-$F814W), are added to the calculated MSRL. These curves are loci of populations with $Y$=0.26--0.37. In panel $a$ of Fig.\ref{F5}, we show some examples ($Y$=0.25,0.29,0.33,0.37) of these loci. We then generate synthetic stellar populations using the technology of artificial stars (ASs). For each population, we generate 2 million ASs in the corresponding magnitude range following a Kroupa-like mass function. We totally generated 2.6$\times10^7$ ASs. These ASs are produced using the appropriate PSF model. We perform the same PSF photometry on these ASs. In order not to dramatically increase the crowding. We repeat this procedure 260,000 times and each time we only input 100 ASs. The recovered ASs thus mimic a simulated observation with the same photometric uncertainty (including noise added by cosmic rays, hot pixels, crowding effect and other artifacts) to real stars. All ASs are homogeneously distributed in the FoV of the observation. The artificial stellar catalog was further reduced using the same procedures applied to real stars. Based on ASs, we obtain the average completeness (the number ratio between the recovered ASs after data reduction and the input ASs) for stars in the cluster and referenced field regions ($\sim$65.8\% and $\sim$82.6\%). 

We finally have 13 artificial stellar populations with $Y$=0.25--0.37, each synthetic population is an SSP, which contains 2 million ASs with realistic photometric uncertainty like the real observation. Because ASs with a flat distribution suffers less crowding effect than the observation, if we directly use the whole sample of ASs, we will underestimate the MS width of ASs, thus overestimating the resulting helium abundance. Because of this, we only select ASs in the cluster region to generate synthetic models. Because we have applied the same data reduction to the artificial stellar catalog, they also share the similar crowding like the observation. As a result, the selected AS samples have a similar spatial distribution to the observation. From each artificial stellar population, we select a subsample with the same luminosity function and the total number of stars as the real observation as a representation. A synthetic MP is a composition of these synthetic populations. In this work, a series of synthetic MPs with different internal helium spreads and fractions of He-rich stars will be used for quantitative comparison, to determine the best-fitting property of stellar populations for the observation. As an example, in panels $b$,$c$ and $d$ of Fig.\ref{F5}, we show the observed MS, a synthetic SSP with $Y$=0.25, and a synthetic MPs with $Y$ ranges from 0.25 to 0.37, where each population with a certain He abundance accounts for 1/13 of the total star number. Simply for a glance, we can see that the observed MS is indeed wider than the synthetic SSP. Its morphology is more consistent with the example MPs which has a helium dispersion of $\Delta{Y}=0.12$ (in this toy model, we have assumed a flat distribution of $Y$). Since a visual examination is unreliable, and a flat distribution of $\Delta{Y}$ is possibly unphysical, in the next Section, we quantify the similarity between models with different $\delta{Y}$ and $\Delta{Y}$ distributions, and the observation using statistical methods.

\section{Main Results} \label{S3}
\subsection{Helium spread among dwarfs}
\begin{figure}[!ht]
\includegraphics[width=\columnwidth]{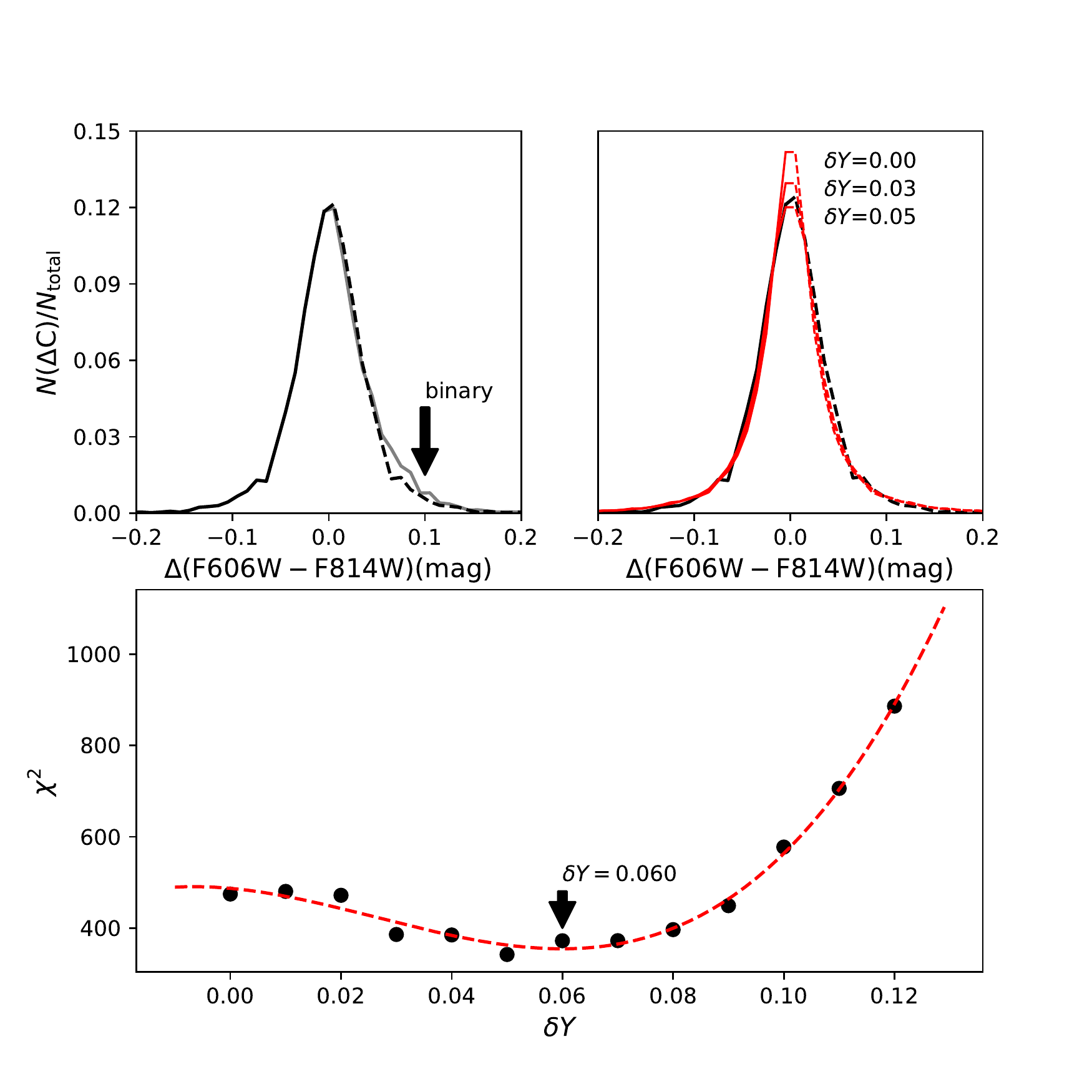}
\caption{Top-left: the observed $\Delta({\rm F606W}-{\rm F814W})$ distribution (grey curve). The distribution for stars with $\Delta({\rm F606W}-{\rm F814W})<$0 mag is indicated by the black curve. The black dashed line represents the mirror of the black solid line (relative to the $\Delta({\rm F606W}-{\rm F814W})=$0 mag). A clear excess of stars on the red side of the MSRL appears, which indicates the contribution of unresolved binaries. Top-right: the same as the top-left panel, with similar distributions of synthetic MPs (red solid/dashed curves). Bottom panel: the $\chi^2$ distribution as a function of $\Delta{Y}$ and the cubic fitting. The implicated $\Delta{Y}$ for minimum $\chi^2$ is indicated by arrow.}
\label{F6}
\end{figure}

\begin{figure}[!ht]
\includegraphics[width=\columnwidth]{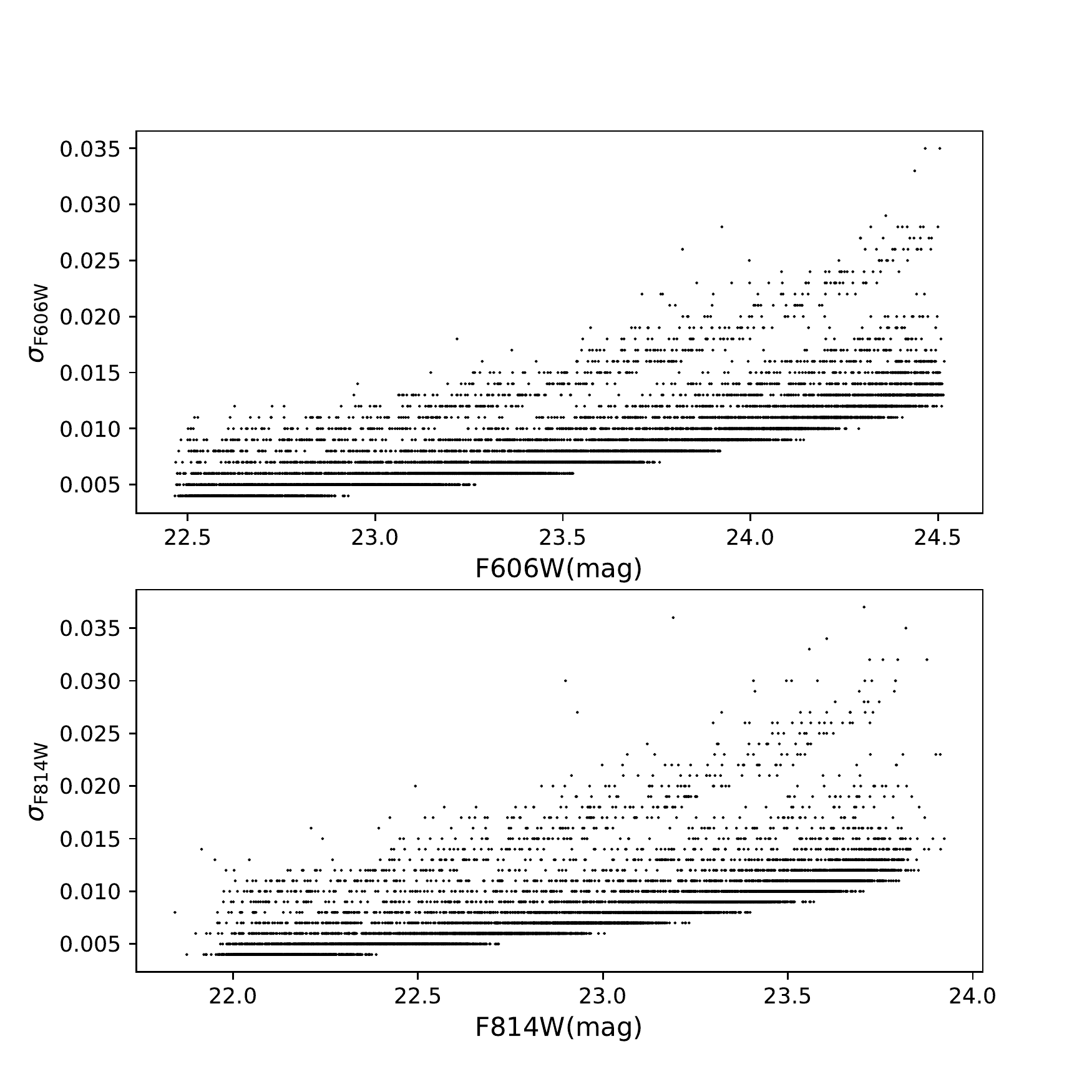}
\caption{The photometric error curves for F606W (top) and F814W (bottom) passbands.}
\label{F7}
\end{figure}

The minimum angular resolution of the {\it HST} at the F814W passband is $\sim$0.1 arcsec, corresponding to $\sim$5000 AU at the distance of the LMC, which is larger than the separation of the widest binaries in the solar neighbourhood \citep{Duqu91a}. We can assume that all binaries in NGC 2210 are unresolved. In the CMD, unresolved binaries will populate a brighter and redder envelope to the MS which can be statistically estimated \citep[e.g.,][]{Milo12a,Li13a}. Although our observed sample must contain some unresolved binaries, however, we do not find any significant unresolved binary feature from the CMD (see Fig.\ref{F4}), which is different from what we found for NGC 1846 \citep{Li21b}. We find that it is difficult to define an appropriate unresolved binary region like what we have done for NGC 1846. The morphology of the MS, particularly its red side, is strongly affected by their binary properties (the fraction, mass-ratio distribution) and line-of-sight blending caused by crowding. These effects hamper an accurate estimation of helium spread. Fortunately, the color of the He-enriched population will be bluer than the bulk population, this behavior is opposite to unresolved binaries and blending. For each star, we calculate their relative color deviation to the MSRL, $\Delta({\rm F606W}-{\rm F814W})$. We have taken a brief visual comparison between the color distributions of two SSPs with/without binaries. Although the color distributions of these two stellar populations are very different in the red sides of their MSs, their blue sides are similar. We thus decide not to analyze stars lying in the red direction of the MSRL, i.e., we only analyze stars with $\Delta({\rm F606W}-{\rm F814W})<$0 mag. In the top-left panel of Fig.\ref{F6}, we show the $\Delta({\rm F606W}-{\rm F814W})$ distribution for all MS dwarfs. We find that the $\Delta({\rm F606W}-{\rm F814W})$ distribution is not symmetric. The standard deviation of the color difference is $\sigma_{\rm color}$=0.0323 mag, while the mean and median photometric errors are $\bar\sigma_{\rm F606W}\approx0.0086$ mag and $\bar\sigma_{\rm F814W}\approx0.0084$ mag, respectively. Their median errors are both 0.008 mag, with 97\% measurements have their photometric errors of $\sigma_{\rm F606W}\leq0.016$ mag and 
$\sigma_{\rm F814W}\leq0.014$ mag, respectively. (see photometric error curves in Fig.\ref{F7}). Clearly, photometric uncertainty cannot explain the observed color spread of the MS solely. We also find a clear excess of `red stars' with $\Delta({\rm F606W}-{\rm F814W})>$0 mag. We determine the fraction of the excess of the `red stars' is $\sim$2.3\%, which is the fraction of unresolved binaries (equal to a mass-ratio of $q=M_1/M_2\gtrsim0.7$) and occasionally blending stars in the line-of-light direction. The total binary fraction, if assuming a flat mass-ratio distribution is $\sim$7.7\%. If assuming a power-low mass-ratio distribution, is $\sim$12.0\%\citep[e.g.,][]{Li13a}. Both are comparable to those for Galactic GCs \citep[5\%--30\%, with a flat mass-ratio distribution][]{Milo12a}, but lower than younger LMC clusters \citep[$\geq50\%$,][]{Li13a,Li21b}.

However, we emphasize that assuming no unresolved binary system would have a negative color deviation is not true. Photometric uncertainty would scatter some unresolved binaries (particularly those with low mass-ratios) to the blue side of the MSRL, which is unavoidable in our analysis. In addition, there must be some (low mass-ratio) binaries belonging to He-rich stellar populations (if present) that are bluer than the MSRL. We think that the number of He-rich binaries must be small. Because most primordial scenarios assume the secondary stellar populations (thus He-rich) form in a more centrally concentrated state than normal stars, which would lead to a more severe dynamical disruption \citep{Hong16a}. Indeed, observations have shown that 2P stars of various clusters have less binary fraction than 1P stars \citep[][but see \cite{Milo20b}]{DOra10a,Luca15a}. In this work we can only minimize (rather than exclude) the binary contamination. 

Using the same method described above, we analyze the relative color distributions for synthetic MPs with different internal helium spreads. We assume a flat $\delta{Y}$ distribution for all these MPs. For example, the model with $\delta{Y}$=0.03 which contains four populations, $Y$=0.25,0.26,0.27,0.28, would have each population occupy 25\% number of stars. For both the observation and the synthetic MPs, we only study their distributions of stars with $\Delta({\rm F606W}-{\rm F814W})<$0 mag. The total number of these stars is $\sim$5000. We divide these stars into 20 color bins, and the bin width is 0.005 mag. This bin size allows us to study their helium distribution in more detail, and in each bin the number of stars is high enough so that they are not strongly affected by statistical variations. To obtain a preliminary comparison, we first analyzed the standard deviation of the color distribution for an SSP, assuming that their color distribution is Gaussian-like. Our result yields $\sigma_{\rm color}$=0.0257 mag. A helium spread of $\delta{Y}$=0.06 ($\sigma_{\rm color}$=0.0317 mag) is required to meet the observation ($\sigma_{\rm color}$=0.0323 mag). 

We then use a $\chi^2$ minimization method to quantify the similarity between models and the observation,
\begin{equation}
\chi^2=\sum_i\frac{(N^{\rm obs}_{\rm i}-N^{\rm mod}_{\rm i})}{N^{\rm obs}_{\rm i}}
\end{equation}
\begin{equation}
N^{\rm obs}_{\rm i}={N^{\rm c}_{\rm i}}-f\frac{A^{\rm c}}{A^{\rm f}}N^{\rm f}_{\rm i}
\end{equation}
where $N^{\rm c}_{\rm i}$ and $N^{\rm f}_{\rm i}$ are the number of stars with their relative colors dropped in the $i$-th bin. The subscript of `c' and `f' means these stars are in the cluster and referenced field regions, respectively. In this work, the area of the cluster region is about 55.4\% of the referenced field region, which denotes $f$=0.554. $A^{\rm c}$ and $A^{\rm f}$ are 0.658 and 0.826, which are the average completeness for cluster and referenced field stars. $N^{\rm obs}_{\rm i}$ is thus the expected number of stars observed in the cluster region, and with their relative colors belong to the $i$-th bin. $N^{\rm mod}_{\rm i}$ is the corresponding number of ASs in the model used for comparison. Finally, we want to examine if the result $\chi^2$ correlates with the model internal helium spread, $\Delta{Y}$. We plot their correlation in the bottom panel of Fig.\ref{F6}. 

We find that the $\chi^2$ distribution exhibits a smooth trend with $\delta{Y}$. The minimum $\chi^2$ occurs at $\delta{Y}=$0.05, with a $\chi^2$=341. To avoid the effect of statistical noise, we used a cubic curve to fit the $\chi^2$--$\delta{Y}$ correlation, which yields a local minimum $\chi^2$ at $\delta{Y}=0.06$. In the top-right panel of Fig.\ref{F6}, we exhibit comparisons between some models and the observation. Indeed, an SSP ($\delta{Y}$=0.0) does not produce the observed $\Delta({\rm F606W}-{\rm F814W})$ distribution, while a MPs with $\delta{Y}$=0.05 exhibit a much better fitting to the observation. For a better illustration, we have symmetrized the $\Delta({\rm F606W}-{\rm F814W})$ distribution although we only analyze stars with $\Delta({\rm F606W}-{\rm F814W})<$0 mag (in Fig.\ref{F7}). 

\begin{figure}[!ht]
\includegraphics[width=1\columnwidth]{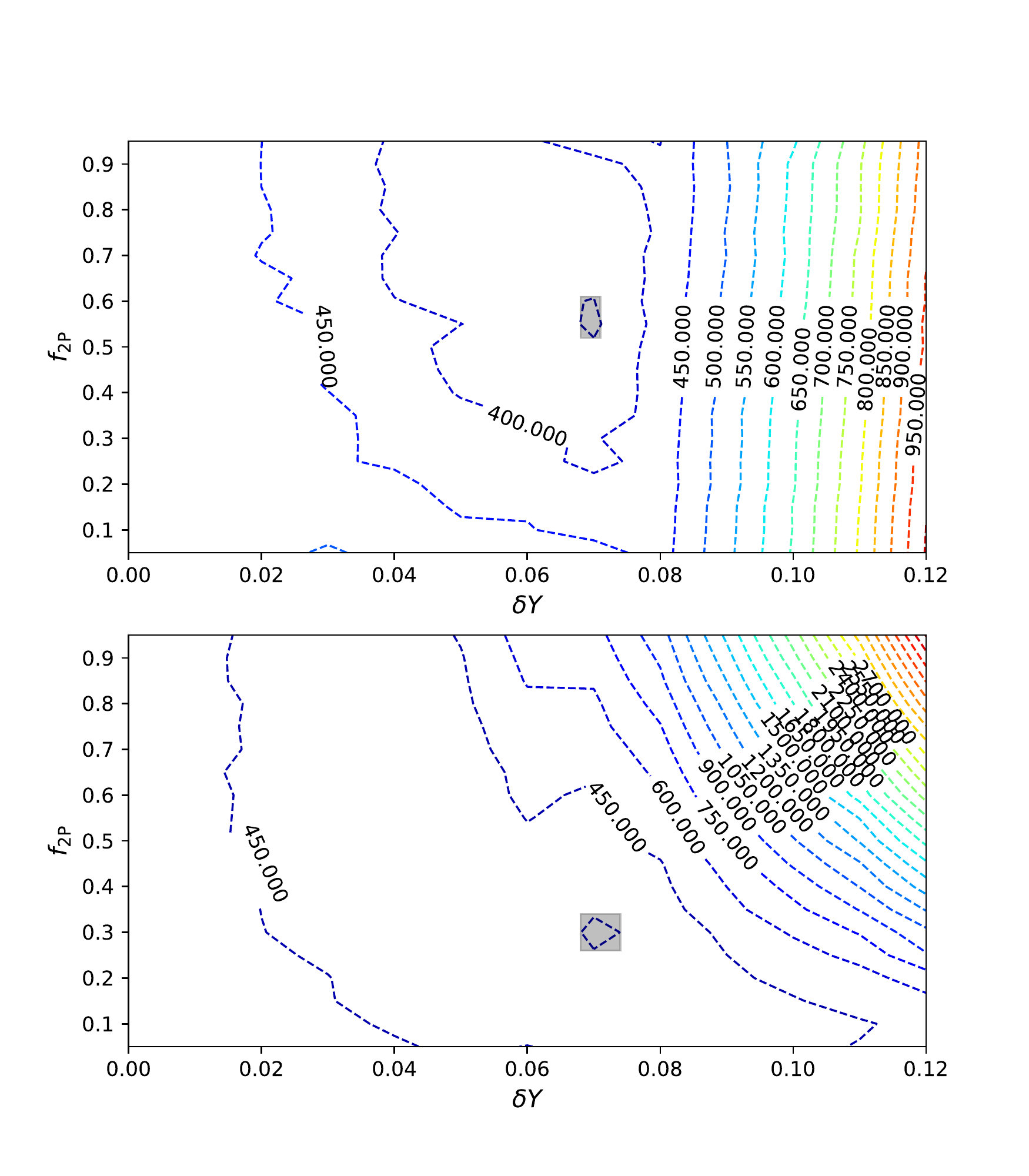}
\caption{The contour of $\chi^2$ distribution as a function of helium spread ($\delta{Y}$) and fractions of 2P stars ($f_{\rm 2P}$). In the top panel, model MPs have a continuous $\delta{Y}$ distribution. In the bottom panel, model MPs have a bimodal $\delta{Y}$ distribution.}
\label{F8}
\end{figure}

The analysis indicates that NGC 2210 may indeed harbor He-rich population stars. A disadvantage is that under the assumption of a flat $Y$ distribution, all these toy models indicate that He-rich stars dominate the sample, which is unrealistic. To derive a more realistic helium distribution, we have generated a series of synthetic MPs with different $\delta{Y}$ and fractions of 2P stars (stars with $\Delta{Y}>0$), $f_{\rm 2P}$. Among the 2P stars, their helium distribution, $\delta{Y}$, is flat. For example, MPs with $\delta{Y}$=0.02 and $f_{\rm 2P}$=30\% would have 70\% normal stars, and each He-rich population ($\delta{Y}$=0.01, 0.02) occupy a number fraction of 15\%. As a result, we totally generated 229 models, including one SSP model ($\delta{Y}$=0), and 228 MPs models with $\delta{Y}$=0.01--0.12 (in step of 0.01), and $f_{\rm 2P}=$5\%--95\% (in step of 5\%). For each model, we calculate its corresponding $\chi^2$. Finally we obtained a 2D-distribution of $\chi^2$ as a function of $\delta{Y}$ and $f_{\rm 2P}$, we plot a contour of the $\chi^2$ distribution in Fig.\ref{F8} (top panel). We find that if the fraction of 2P stars is too low ($f_{\rm 2P}\leq20\%$), it yields a $\delta{Y}\sim0.08$, but the $\chi^2$--$\delta{Y}$ distribution is very noisy. For $f_{\rm 2P}\geq40\%$, the $\chi^2$--$\delta{Y}$ correlation becomes smooth. They all report a best-fitting $\delta{Y}$ ranges from 0.06--0.10. The minimum range of $\chi^2$ occurs at $\delta{Y}\sim$0.068--0.071 and $f_{\rm 2P}\sim$52\%--61\%(shadow region in the top panel Fig.\ref{F8}), corresponding to a $\chi^2\leq$364.0, within this region the variation of $\chi^2$ is dominated by noise. In summary, if we assume a continuous $\delta{Y}$ distribution, NGC 2210 is likely to have $\sim55\%$ He-rich stars, with an internal helium spread of $\delta{Y}=0.069^{+0.002}_{-0.001}$.

However, such a high fraction of 2P stars is surprising. According to primordial scenarios, the primordial population (1P) stars form earlier than the chemically enriched population (2P) stars with a more extended configuration. As a result, the 1P stars are more easily stripped by the external galactic tidal field \citep[e.g.,][]{Derc08a}. The number ratio between the 2P and 1P stars would be lower for GCs in a weaker external tidal field (the LMC) than those in the Galaxy. A possible explanation is that the $\delta{Y}$ distribution is discrete (such as NGC 2808) rather than continuous. We therefore generated another set of models, in these models, MPs have a bimodal distribution of $\delta{Y}$. For example, a model with 30\% He-rich stars and $\delta{Y}$=0.02 only contains two populations, i.e., 70\% normal stars ($\delta{Y}$=0.00) and 30\% He-rich stars ($\delta{Y}$=0.02). Again, under the adoption of bimodal distributions for MPs, we plot the $\chi^2$ distribution as a function of $\delta{Y}$ and $f_{\rm 2P}$, which is present in the bottom panel of Fig.\ref{F8}. This time, the lowest $\chi^2$ region occurs at $\delta{Y}$=0.068--0.074 and $f_{\rm 2P}$=26\%--34\% (shadow region in the bottom panel Fig.\ref{F8}). We find that MPs with a bimodal distribution of $\delta{Y}$ can better reproduce the observation, as they return a lower $\chi^2$ ($\lesssim$332) than the case of a flat $\delta{Y}$ distribution ($\lesssim364$). We suggest that a small fraction of 2P stars is more reasonable. Indeed, studies have shown that chemically enriched populations in some LMC clusters only occupy a small fraction of 10\%--20\% \citep{Holl19a,Dond21a}. We conclude that NGC 2210 most likely harbors $\sim30\%$ He-enriched stars, with a maximum helium spread of $\delta{Y}=0.071\pm0.003$.

Our results indicate that NGC 2210 is very different to NGC 1846, the latter is likely an SSP cluster ($\delta{Y}<$0.02). However, the detection of a helium variation does not indicate that the He-rich stars do belong to the 2P. In addition to He-enhancement, if we strictly define the 2P stars are those with  Na, C, N, O variations, some 1P stars (stars without these specific chemical patterns) are found to have helium spread \citep[e.g.,][]{Milo17a} as well, although the reason remains unclear yet. How to determine if the derived helium spread is an internal spread of 1P stars or if it indicates the presence of MPs? One way is to examine their radial distributions. If both He-normal and He-rich stars are 1P stars, they should be fully mixed in spatial. Otherwise 1P and 2P stars may exhibit different central concentrations, according to primordial scenarios. In this work, we cannot determine if a specific star is He-enriched or normal. Alternatively, we compared each star's color deviation to their photometric uncertainty. Stars with a color deviation $|\Delta({\rm F606W-F814W})|$ larger than three times the expected color uncertainty are defined as He-rich stars. Using this criterion, 33.7\% stars (1547 of 4557) are defined as He-rich stars. This is  consistent with the indication derived by bimodal $\delta{Y}$ distribution models ($\sim$30\%). If the observed MS dwarfs are SSP, we would expect only $14\pm4$ stars (0.3\%) meet this criterion. 

\begin{figure}[!ht]
\includegraphics[width=1\columnwidth]{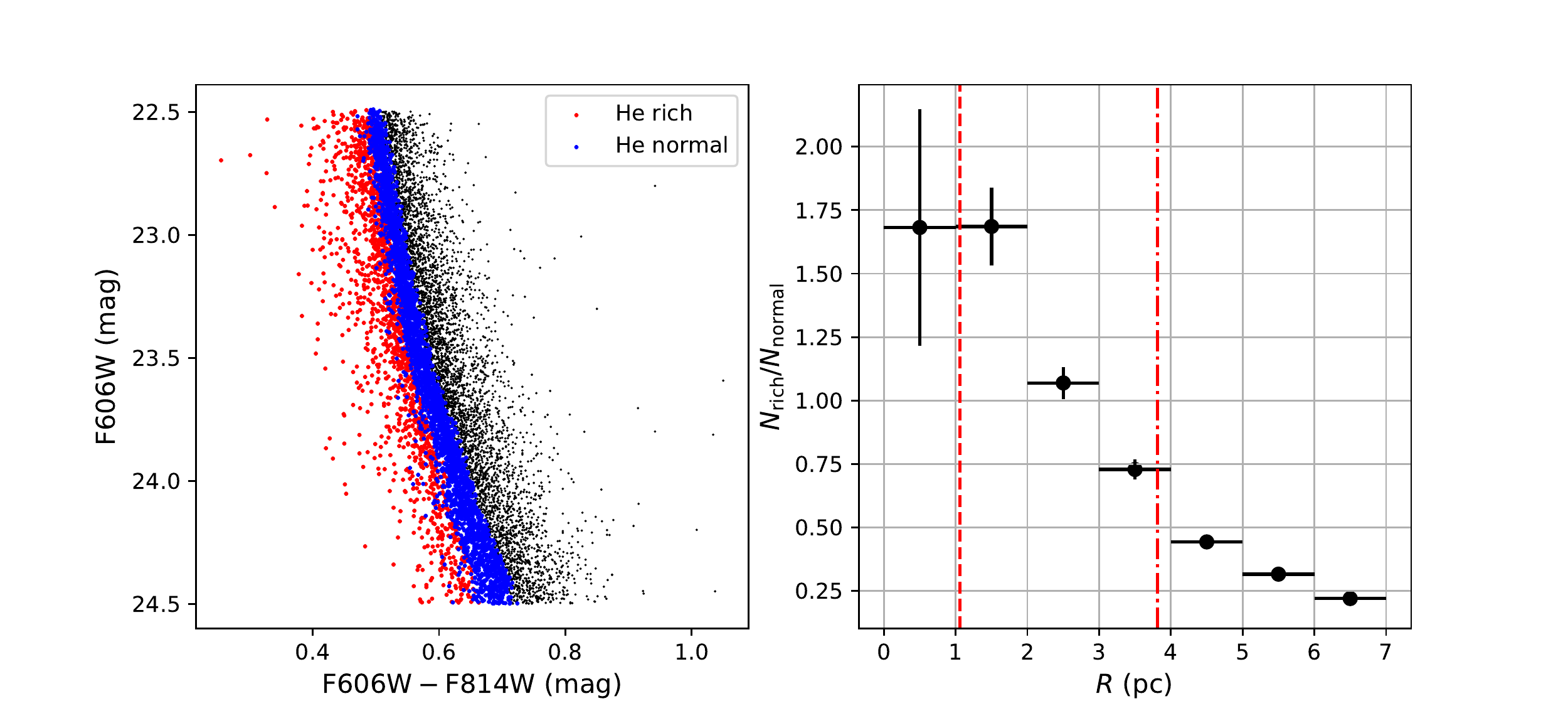}
\caption{Left, the CMD of He-rich (red dots) and normal (blue dots) stars. Right, the number ratio radial distribution between He-rich and normal stars. The red dashed and dash-dotted lines represent the positions of the core and half-mass radius, respectively. Associated error bars are Poisson-like.}
\label{F9}
\end{figure}

We divide both the normal and He-rich stars into seven radial bins ranging from the cluster center to a radius of 7 pc ($\sim$2$r_{\rm hm}$), with a bin size of 1 pc, the latter is roughly the core size ($r_{\rm c}$) of NGC 2210 \citep{Ferr19a}. We study the radial profile of the number ratio between He-rich and normal stars. If He-rich stars belong to a secondary stellar population formed in the cluster center after the formation of 1P stars, the number ratio radial profile should exhibit a decreasing trend from the cluster center to its outskirt. In the right panel of Fig.\ref{F9} we exhibit this number ratio radial profile. We cannot tell any radial difference between the He-rich and normal stars within 2$r_{\rm c}$. We find a significant decreasing trend in the range of 2--7 pc, indicating that the He-rich population has a more compact configuration than normal stars. Since He-rich and normal stars are all in the same magnitude range, this radial difference cannot be explained by their completeness difference in the radial direction. In addition, He-rich dwarfs are less massive than normal dwarfs with similar luminosities, which further strengthens the implication that they must be initially much more compact than normal stars. 


\begin{figure}[!ht]
\includegraphics[width=1\columnwidth]{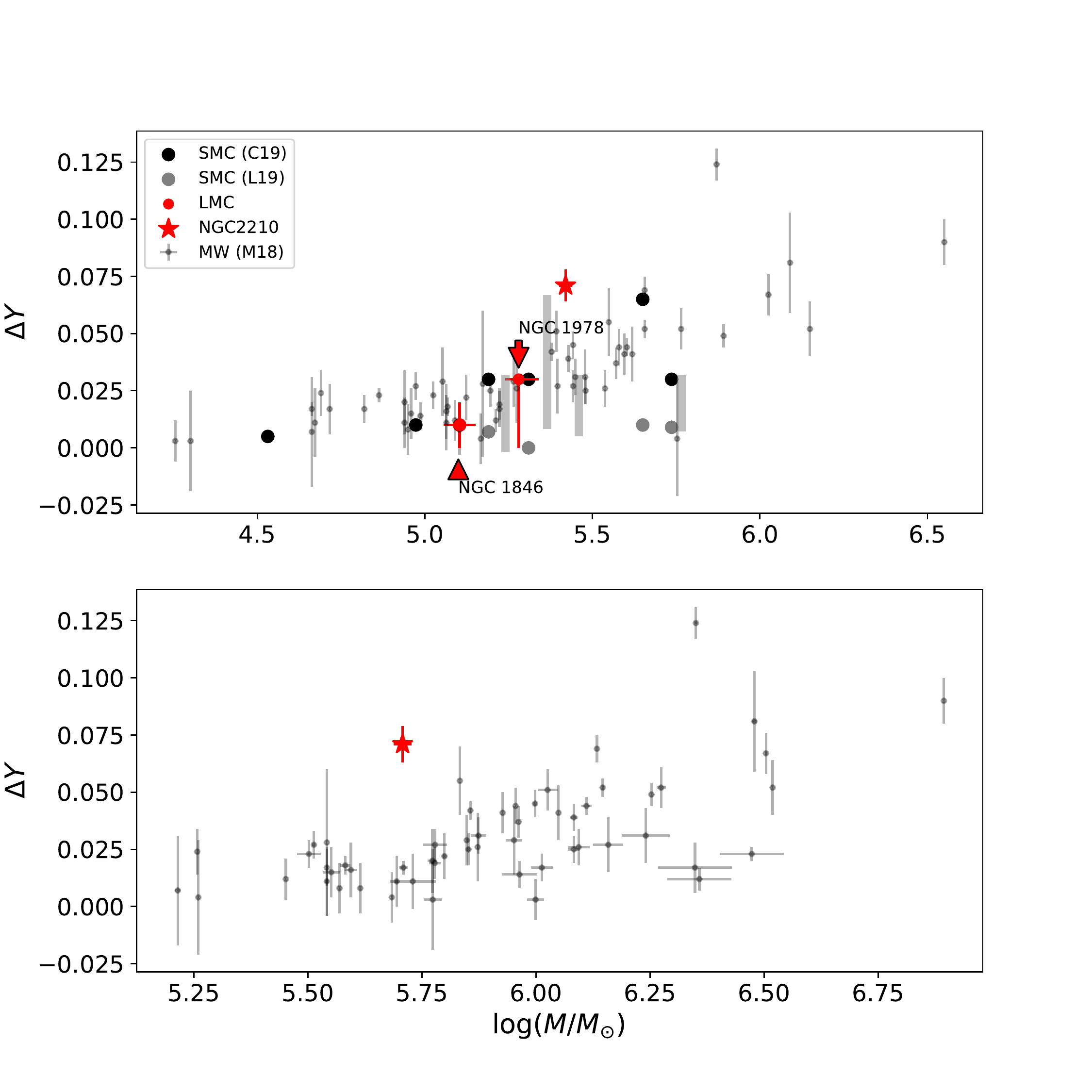}
\caption{Top: the internal helium spreads and the clusters present-day masses diagram, $\delta{Y}$--$\log(M/M_{\odot})$. Red circles and a pentagram (NGC 2210, this work) are LMC clusters. Small grey dots are Milky Way GCs. Dark/light grey circles are SMC clusters. For NGC 1846 we have used its expected total mass at $\sim$10 Gyr. Bottom: the correlation between the internal helium spreads and the clusters initial masses, for Milky Way GCs and NGC 2210.}
\label{F10}
\end{figure}

\cite{Milo17a} have derived a clear correlation between the internal maximum helium spread and cluster's present-day mass. We want to examine if NGC 2210 also follows the trend. We compare our results with Galactic GCs \citep{Milo17a}, LMC clusters \citep{Li21b,Ji22a}, other SMC clusters with internal helium spread studied in literatures \citep[][hereafter C19/L19]{Chan19a,Lagi19a}. This result is shown in the top panel of Fig.\ref{F10}. We find that although the helium spread of NGC 2210 is relatively higher than its Galactic counterparts, it is consistent with the same correlation. If we only consider LMC clusters, this correlation is likely steeper than that for Galactic GCs. 

The cluster initial mass should be a more appropriate parameter that decide the property of MPs than their present-day masses. For Galactic GCs, \cite{Baum19a} have integrated their orbits backward in time to derive the cluster initial masses, taking the effects of dynamical friction and mass-loss of stars into consideration. Using the same method, we have calculated the initial mass of NGC 2210 using N-body simulations, which yields 5.1$\times10^5$ $M_{\odot}$. We find this cluster have lost very limited mass through dynamical effects because of its high mass and the weak tidal field of the LMC. The difference between the present-day and the initial mass is almost entirely due to stellar evolution mass loss. As a result, the present-day number ratio between 1P and 2P stars of NGC 2210 should be almost identical to its initial value. In the bottom panel, we present the helium spread--initial mass relationship for Galactic GCs (grey dots) and NGC 2210 (the red star). It turns out that NGC 2210 indeed harbor a higher internal helium spread than its Galactic counterparts with similar initial masses. It remains unclear if this would indicate that LMC GCs would exhibit a different helium spread--initial mass correlation, studies of more LMC samples are required. The initial masses for LMC/SMC clusters will be present in a forthcoming article. 

\subsection{Comparing with evolved giants}
\begin{figure}[!ht]
\includegraphics[width=1\columnwidth]{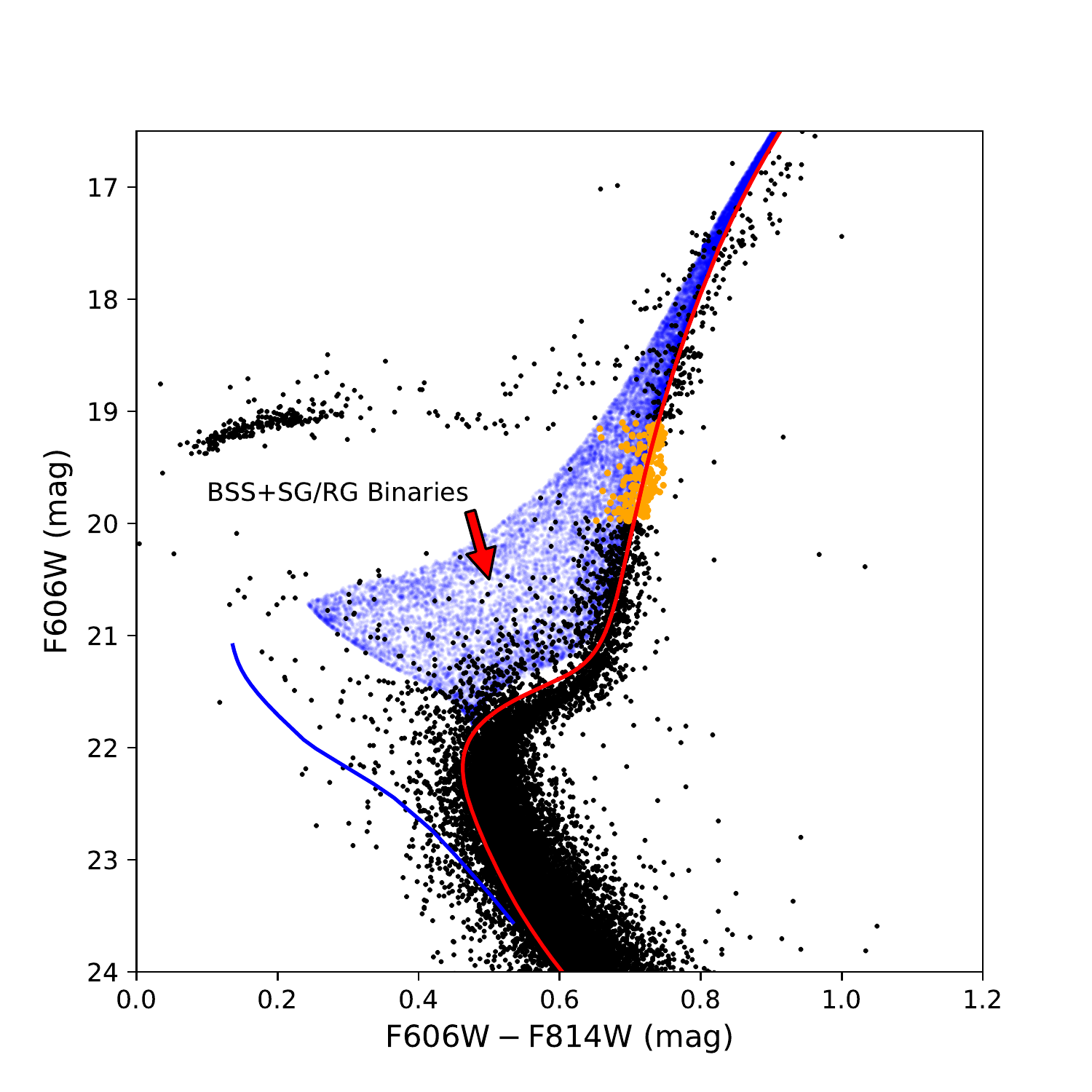}
\caption{RG stars (orange dots) were selected to study the helium spread. The blue dots are simulated BSS-SG/RG binaries (unresolved). The solid blue curve describes the BSS locus. The red curve is the best-fitting isochrone. }
\label{F11}
\end{figure}

Using the same method, we have derived the helium spread among red-giant stars of NGC 2210. The sample we used is red-giant (RG) stars lying significantly above the bottom of the RGB (F606W$\sim$21.28 mag) and below the RGB bump (F606W$\sim$17.56 mag), in the range of 19.08 mag$\leq$F606W$\leq$19.98 mag. We constrain the sample RG stars with a color range of 0.65 mag$\leq$F606W$-$F814W$\leq$0.75 mag. The selections of the magnitude and color ranges are arbitrary. We confirm that the calculated ridge line of the selected RGB part is close to the best-fitting isochrone. We exhibit the selected RG stars and the best-fitting isochrone in Fig.\ref{F11}. However, NGC 2210 exhibits many blue straggler stars (BSSs).  A significant fraction of these BSSs may lie in a mass-transferring binary system, where another binary component is likely a sub-giant (SG) or a RG star. These BSS-SG/RG binaries would be distributed in a region between the RGB and the BSS locus, partially overlapping with the He-rich RGB. We generate a large number of artificial BSS-SG/RG binaries and plot them in the CMD of the cluster. For each BSS-SG/RG binary, the BSS is randomly selected from the BSS locus, which is described by a 1 Gyr-old isochrone (with other parameters identical to the best-fitting isochrone). The SG/RG star is selected from the best-fitting isochrone. We find that the region of the BSS-SG/RG binaries exhibits a clear top boundary, which gradually decreases from the region close to the TP-AGB toward the TO region of the BSS. Between this boundary and the HB region, there are no stars. This ideally describes the observation. Given that single stars in the Hertzsprung gap region evolve rapidly, we conclude that most observed stars in this region are unresolved BSS-SG/RG binaries. Some of these binaries will strongly contaminate the He-rich RG population (if the RG component dominates the flux of the binary system). Because of this, we expect that the helium spread derived from the width of the RGB would be overestimated if we cannot rule out the binary contamination. 

\begin{figure}[!ht]
\includegraphics[width=1\columnwidth]{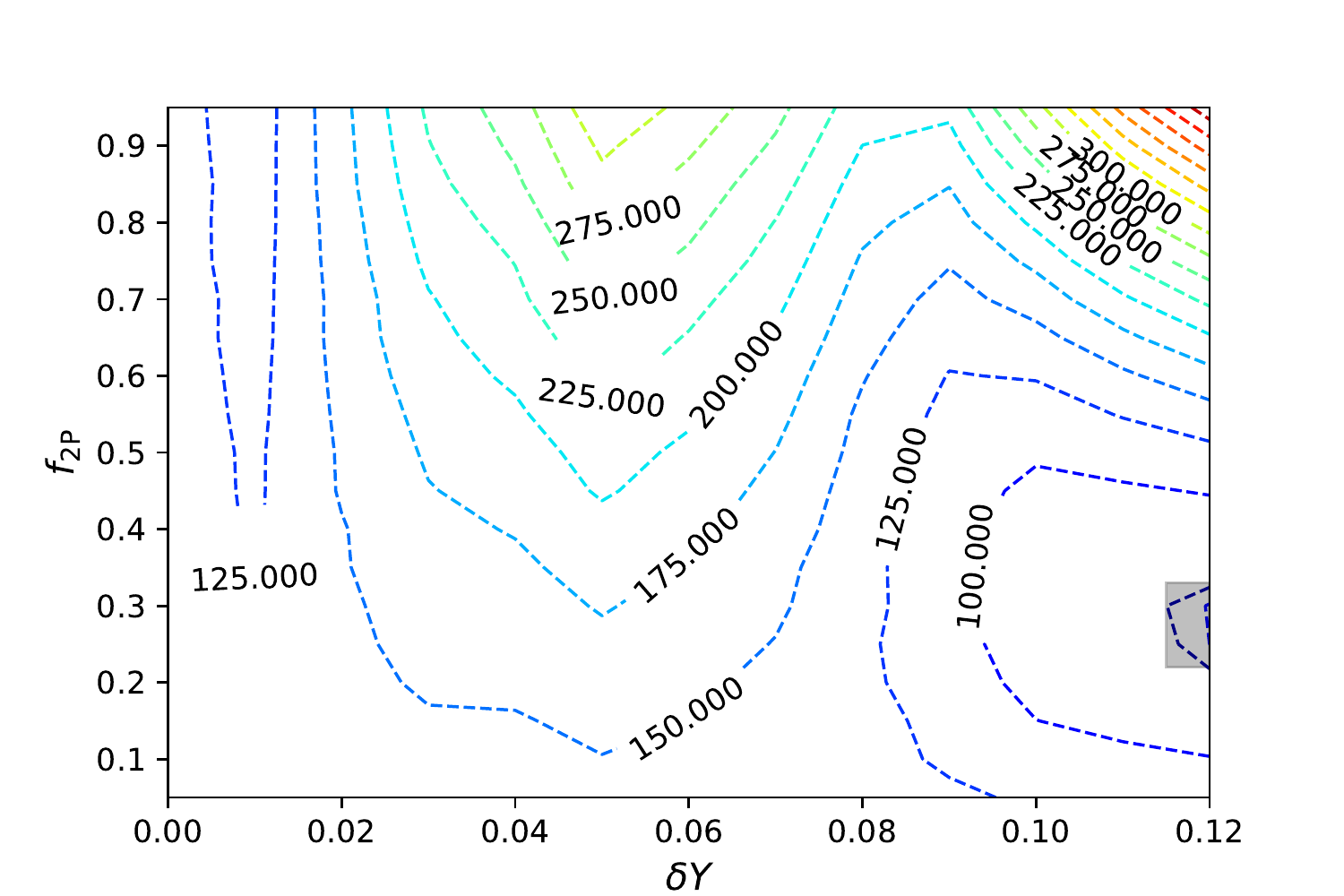}
\caption{The same as Fig.\ref{F8}, but for RG stars. The helium distribution of the model RGB is bimodal.}
\label{F12}
\end{figure}

Indeed, our analysis report that, if we assume that helium spread fully accounts for the width of the RGB, the internal helium spread will reach $\delta{Y}$=0.12, the upper limit of the model we used. Binaries play an important role in the broadening of the RGB. The fraction of He-rich stars is 22\%--33\%, which is consistent with the result derived from MS stars (Fig.\ref{F12})

\begin{figure}[!ht]
\includegraphics[width=1\columnwidth]{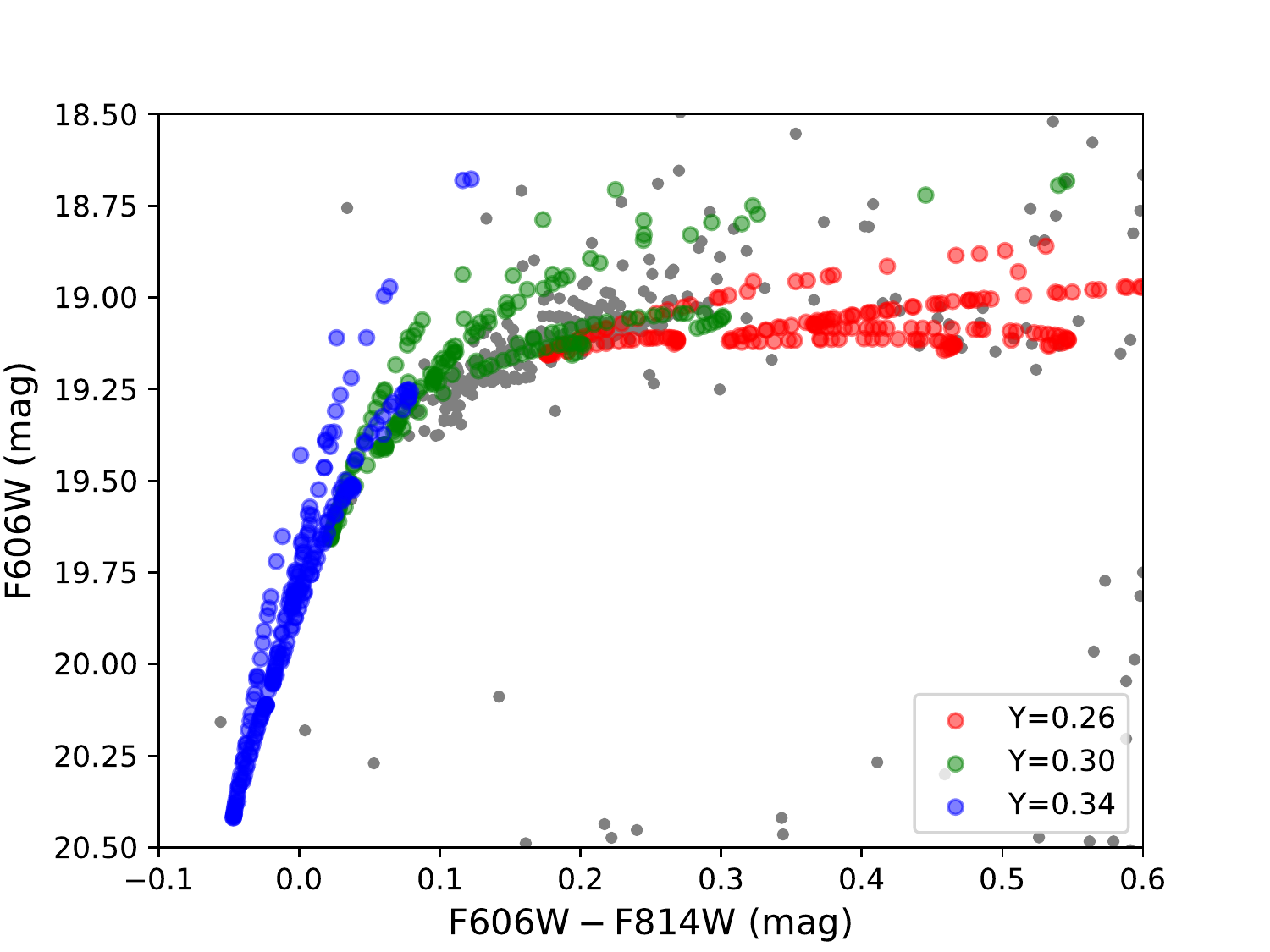}
\caption{The CMD of the HB, overlap with the simulated HB populations with different helium abundances and mass loss rates.}
\label{F13}
\end{figure}

Because PGPUC does not calculate HB phase, we used the Modules for Experiments in Stellar Astrophysics \citep[MESA,][]{Paxt11a} to examine if our result also fits the HB morphology. We use the MESA to calculate three 12.5 Gyr-old HB loci with $Y=$0.26, 0.30, 0.34, respectively. They thus briefly describe the morphology of HB with $\delta{Y}\sim0.08$. The adopted metallicity is [Fe/H]=$-$1.92 dex (the same as the best-fitting PGPUC isochrone). The most important parameter controlling the HB morphology is the mass loss rate. During the RGB phase, it is described by Reimer's mass loss rate \citep[$\eta_R$,][]{Reim75a,Reim77a}. The mass loss rate for RG stars varies from cluster to cluster, covering a range of $\eta_R<0.2$ to $\eta_R>0.6$ \citep{Tail20a}. Because the mass loss rate in our model is a free parameter, the helium spread among HB stars is uncertain. To make a qualitative comparison, we first conservatively set a $\eta_R=0.2$ to our model. In this case, the simulated HB with $\delta{Y}$=0.08 exhibits a more extended morphology than the observation. Under the adoption of $\eta_R=0.2$, a $\delta{Y}$=0.04 ($Y$=0.26--0.30) is sufficient to explain the length of the observed HB. We then calculated another two model sets with $\eta_R=0.1$ and 0.05. We find that once we adopt $\eta_R=0.05$, the simulated HB population with a helium spread of $\delta{Y}=0.04-0.08$ fits the observation better. We finally constructed a stellar HB population with three helium abundances ($Y=$0.26, 0.30, 0.34), with each sub-population containing stars with different mass loss rate $\eta_R=0.05-0.20$. We present the fitting to the observation in Fig.\ref{F13}.

\section{Discussion and Conclusion} \label{S4}
Before we discuss the physical implications of our results, we first examine if there is no helium spread, at what value an additional differential reddening is required to fully account for the width of the MS, by comparing the synthetic stellar population with different differential reddening with the de-reddened observation. Our analysis reports an additional differential reddening of $\sigma_{E(B-V)}=0.004$ mag is required to explain the observed MS. This is about 20 times the expected differential reddening residual. We confirmed that the signature of such an additional reddening is significantly enough to be revealed through our de-reddening method, if present. Therefore, we conclude that the broadening of the width cannot be fully explained by differential reddening.

Another effect that would contribute to the width of the MS is metallicity spread. Decreasing the metallicity would reduce the stellar atmospherical opacity, leading to a decrease in cooling efficiency, thus, an increase in the stellar surface temperature. For this reason, stars with lower metallicity will look bluer than normal stars at each evolutionary stage, populating a bluer MS. We then generate a series of isochrones with different metallicities, and compare their loci with the $Y=0.33$ isochrone, the latter corresponds to the $\Delta{Y}=0.07$ stellar population locus. However, we find that even if we generate an isochrone with $Z$=0.00001 (the lower metallicity limit of the PGPUC model), the color difference between this isochrone and the best-fitting isochrone ($Z$=0.00016) cannot describe the width of the MS. Since we only concern at what value a metallicity spread can describe the width of the MS, we release the upper limit of the metallicity. Although in that case, the isochrone may not be able to describe the CMD well. We find that a metallicity spread from $Z$=0.0002 to $Z$ = 0.001 ([Fe/H]=$-1.92$$\sim$$-1.22$ dex) is required to fully account for the observed width of the MS. Such a metallicity spread would produce a very wide SGB and RGB. The inconsistency between the model isochrones and the observation can be easily derived visually. We thus exclude the presence of a dramatic metallicity spread among NGC 2210 members. We also confirm that a spread of $[\alpha/{\rm Fe}]$=0.00--0.30 dex (the maximum input range allowed for the PGPUC model) has a negligible contribution to the width of the MS.

A similar analysis reports that RGB stars may contain $\sim$30\% He-rich stars, which agrees with our analysis for MS dwarfs. But it implies a higher helium spread content, $\delta{Y}$--0.12. As we have illustrated, this is likely due to the contamination of BSS/RG binaries. The morphology of the HB, is very short, however. Our simulation indicates that under the fixed mass loss rate of $\eta_R=0.2$, only a helium spread of $\delta{Y}$=0.04 is sufficient to explain the observed HB, which is lower than the value derived from MS dwarfs ($\delta{Y}=0.06-0.07$). A lower mass loss rate down to $\eta_R=0.05$ would fit the observation better with a higher helium spread up to $\delta{Y}$=0.08. However, this would indicate that RGB stars in NGC 2210 experienced less mass loss than Galactic GCs \citep{Tail20a}. Speculation is that the parameter that controls the HB color extension is environment-dependent. The tidal fields of their host galaxies affect the mass loss during their RGB phase. We highlight that another LMC GC, Hodge 11, exhibits an extended HB \citep[][their figure 19]{Gill19a}. An accurate determination of internal helium variation for this cluster would be crucial.

Since analyses of RGB and HB members are affected by binaries and mass loss rates from star to star, both are very uncertain. We are now back to discussing the results derived from MS dwarfs, as the helium spread inferred by MS dwarfs is indicative of the helium contents of the clouds from which multiple populations formed. In summary, in this work, our main conclusions are,

\begin{itemize}
\item[1.] NGC 2210 does exhibit a helium spread of $\delta{Y}\sim0.06$--0.07. The number ratio of He-rich stars to the whole population is about $\sim$30\%, if assume that the $\delta{Y}$ distribution is bimodal. Otherwise it would be more than half, 55\%, if the $\delta{Y}$ distribution is continuous. 
\item[2.] He-rich stars are more centrally concentrated than normal stars, indicating that the detected helium spread is not an internal spread among 1P stars, but for two stellar populations formed in different initial configurations. 
\item[3.] The internal helium spread, $\delta{Y}$, of NGC 2210 is consistent with the correlation between the helium spread and the clusters' present-day mass for Galactic GCs. If we only consider LMC clusters, this correlation is even steeper. 
\end{itemize}

In a previous study/search on multiple populations in NGC2210, \cite{Gill19a} detected a broadened MS and estimated that the second population includes 20$\pm$5\% of stars. The fraction of second-population stars derived in this work is higher than theirs, if we assume a bimodal helium distribution (26\%--34\%). In addition, we inferred the internal helium variation of NGC2210 by assuming that the MS color broadening is mostly due to an helium scatter.
This is a reasonable hypothesis because most elemental absorption lines concentrated in the UV band, their effects on the color of F606W$-$F814W are negligible \citep[see figure 5 of][as an example]{Milo18a}.  


For LMC clusters, the positive correlation between the internal helium spread, $\Delta{Y}$, and the present-day GC mass \citep[][]{Li21b}, is similar to that for Galactic GCs \citep{Milo18a}. If the $\delta{Y}$ distribution is (close to) bimodal, the 2P fraction of NGC 2210 would be smaller than its Galactic counterparts with comparable masses \citep[][their figure 8]{Milo22a}. This would support scenarios where GCs preferentially lose their 1P stars, as 1P stars of LMC GCs experience weaker tidal stripping than Milky Way clusters. The fact that He-rich stars are more centrally concentrated than He-normal stars in NGC 2210 is in qualitatively agreement with the prediction from the main scenarios on the formation of multiple populations \citep[e.g.,][]{Krau13a,Dant16a,Calu19a,Giel18a,Wang20a}. After the gas expulsion, both 1P and 2P stars escape during the long-term evolution by the galactic tide, and a large amount of time (up to $\sim$20$t_{\rm rh}$) is needed to fully mix the 1P and 2P stars \citep[e.g.,][]{Vesp13a}. According to \cite{McLa05a}, the $t_{\rm rh}$ for NGC 2210 is 1.0--1.2 Gyr ($\log{t_{\rm th}}$=9.01--0.06, model dependent). If the half-mass relaxation timescale does not significantly change during its evolution, the dynamical age of NGC 2210 is 10--12$t_{\rm rh}$. At this dynamical age, we would expect that at least the part with $r\leq{r_{\rm rh}}$ will already be fully mixed, which is inconsistent with our observation. We speculate that this is because \cite{McLa05a} used a simplified model to calculate the $r_{\rm rh}$ in which the average stellar mass in NGC 2210 is assumed $M_{\star}$=0.5$M_{\odot}$, which introduces an additional uncertainty. Using the same method of \cite{Baum19a}, our calculation yields a longer half-mass relaxation time of $t_{\rm rh}\sim$3.2 Gyr for NGC 2210, indicating that NGC 2210 is only 3--4 $t_{\rm rh}$ old. This is in good agreement with our observation as only stars within the core radius are fully mixed.

The difference between NGC 2210 and NGC 1846, where the latter exhibit a minimum helium spread, may be controlled by the difference in their masses or ages, as NGC 1846 is younger and, less massive (at the age of $\sim$10 Gyr) than NGC 2210. To determine which parameter plays the critical role, further studies focusing on younger and more massive LMC clusters (i.e., NGC 1850) in terms of their helium distributions are required. Again, this would require high precision photometry focusing on their MS as young clusters usually do not have well-populated RGB and HB. 

If the He-rich stars detected in NGC 2210 represent 2P stars, they may exhibit common patterns of MPs, i.e., Na-O anti-correlation, C-N anti-correlation. A spectroscopic analysis of these stars is not possible because these stars are too faint. Alternatively, utilizing UV-optical photometry may statistically examine whether or not these stars are different in C, N, O abundances \citep[e.g.,][]{Li21a}. Again, this requires deep photometry which will consume lots of {\it HST} time. The next-generation Chinese Space Station Telescope ({\it CSST}) with similar parameters to the {\it HST}, can take over this task with a larger FoV \citep{Li22a}. 

\newpage
\acknowledgements 
{C. L. is supported by the National Key R\&D Program of China (2020YFC2201400). D.J. acknowledge support from the National Natural Science Foundation of China (Nos 12073070, 11733008). C.L. and L.W. acknowledge support from the one-hundred-talent project of Sun Yat-sen University and the National Natural Science Foundation of China through grant 12073090 and 12233013. B.T. gratefully acknowledges support from the National Natural Science Foundation of China under grant No. U1931102, and the Natural Science Foundation of Guangdong Province under grant No. 2022A1515010732.This work is also supported by the China Manned Space Project with NO.CMS-CSST-2021-A08, CMS-CSST-2021-B03, National Key R\&D Program of China with No. 2021YFA1600403 and CAS `Light of West China' Program. Y.W. acknowledges the support by the Special Research Assistant Foundation Project of Chinese Academy of Sciences.}

\end{CJK*}
\end{document}